\documentclass[useAMS,usenatbib,usegraphicx,letterpaper]{mn2e}

\newcommand{\Msun}{\mbox{$\mathrm{M}_{\odot}$}}

\newcommand{\mwd}{\mbox{$M_{\mathrm{WD}}$}}
\newcommand{\twd}{\mbox{$T_{\mathrm{eff}}$}}
\newcommand{\sugr}{\mbox{$\mathrm{log}\,g$}}
\usepackage{amssymb}
\title[Pulsating WDs in WDMS binaries]{Discovery of ZZ\,Cetis in detached white dwarf plus main-sequence binaries}

\author[S.Pyrzas et al.]{
S. Pyrzas$^{1}$\thanks{E-mail: stylianos.pyrzas@gmail.com},
B. T. G\"ansicke$^{2}$,
J. J. Hermes$^{2}$,
C. M. Copperwheat$^{3}$,
A. Rebassa-\newauthor
Mansergas$^{4}$,  
V. S. Dhillon$^{5}$,
S. P. Littlefair$^{5}$,
T. R. Marsh$^{2}$,
S. G. Parsons$^{6}$,\newauthor 
C. D. J. Savoury$^{5}$, 
M. R. Schreiber$^{6}$,
S. C. C. Barros$^{7}$,
J. Bento$^{8}$,
E. Breedt$^{2}$ and 
P. Kerry$^{5}$
\\
$^{1}$Instituto de Astronom\'ia, Universidad Cat\'olica del Norte, Avenida Angamos 0610, Casilla 1280, Antofagasta, Chile \\
$^{2}$Department of Physics, University of Warwick, Coventry, CV4 7AL, UK\\ 
$^{3}$Astrophysics Research Institute, Liverpool John Moores University, Twelve Quays House, Birkenhead CH41 1LD, UK \\
$^{4}$Kavli Institute for Astronomy and Astrophysics, Peking University, Beijing 100871, China \\
$^{5}$Department of Physics and Astronomy, University of Sheffield, Sheffield S3 7RH, UK \\
$^{6}$Departamento de F\'isica y Astronom\'ia, Universidad de Valpara\'iso, Avenida Gran Breta\~na 1111, Valpara\'iso, Chile \\
$^{7}$Aix Marseille Universit\'e, CNRS, LAM, UMR 7326, 13388, Marseille, France \\
$^{8}$Department of Physics and Astronomy, Macquarie University, NSW 2109, Australia \\
}
\begin{document}

\date{Accepted . Received ; in original form}

\pagerange{\pageref{firstpage}--\pageref{lastpage}} \pubyear{2014}

\maketitle

\label{firstpage}

\begin{abstract}
We present the first results of a dedicated search for pulsating white dwarfs (WDs) in detached white dwarf plus main-sequence binaries. Candidate systems were selected from a catalogue of WD+MS binaries, based on the surface gravities and effective temperatures of the WDs. We observed a total of 26 systems using ULTRACAM mounted on ESO's 3.5\,m New Technology Telescope (NTT) at La Silla. Our photometric observations reveal pulsations in seven WDs of our sample, including the first pulsating white dwarf with a main-sequence companion in a post common envelope binary, SDSS\,J1136+0409. Asteroseismology of these new pulsating systems will provide crucial insight into how binary interactions, particularly the common envelope phase, affect the internal structure and evolution of WDs. In addition, our observations have revealed the partially eclipsing nature of one of our targets, SDSS\,J1223-0056.
\end{abstract}

\begin{keywords}
stars: white dwarfs - stars: low-mass - binaries: close - binaries: eclipsing - asteroseismology
\end{keywords}


\section{Introduction}
\label{sec:intro}

\begin{table*}
\setlength{\tabcolsep}{0.75ex}
\centering
\caption{Information on all our target WD+MS binaries. We provide names, coordinates, SDSS $u, g, r$ magnitudes$^{*}$, \twd, \sugr\, and \mwd\, values. The penultimate column gives the classification of each target (see text for details), while the last column gives the length of the observations.}
\label{tab:allinfo}
\begin{tabular}{@{}ccccccccccc@{}}
\hline 
SDSS\,J & RA [J2000] & Dec [J2000] & $u$ & $g$ & $r$ & \twd\,$^{a}$ [$\mathrm{K}$] & \sugr\,$^{b}$ & \mwd\,$^{c}$ [\Msun] & Class & Obs. Length [hrs] \\ 
\hline 
0017-0024 & 00:17:26.63 & -00:24:51.1 & 19.66 & 19.21 & 18.96 & $14594\pm1487$ & $7.96\pm0.30$ & $0.60\pm0.18$ & WDMS & 2.46 \\
0021-1103 & 00:21:57.90 & -11:03:31.6 & 19.22 & 18.57 & 18.10 & $11045\pm164 $ & $8.78\pm0.10$ & $1.08\pm0.05$ & WDMS & 2.89 \\
0052-0051 & 00:52:08.42 & -00:51:34.6 & 19.04 & 18.30 & 17.72 & $12300\pm427 $ & $8.46\pm0.12$ & $0.90\pm0.08$ & WDMS & 6.82$^{1}$ \\
0111+0009 & 01:11:23.89 & +00:09:35.3 & 19.02 & 18.42 & 17.83 & $12321\pm461 $ & $7.50\pm0.25$ & $0.37\pm0.11$ & WDMS & 2.85 \\
0124-0023 & 01:24:03.11 & -00:23:01.1 & 20.29 & 18.88 & 17.75 & $11972\pm1509$ & $6.70\pm0.44$ & $0.16\pm0.10$ & WDMS & 2.02 \\
0203+0040 & 02:03:51.28 & +00:40:25.1 & 20.32 & 19.43 & 18.67 & $10794\pm475 $ & $8.17\pm0.36$ & $0.71\pm0.22$ & WDMS & 2.93 \\
0212+0018 & 02:12:39.45 & +00:18:56.9 & 19.69 & 19.22 & 18.87 & $13852\pm2865$ & $8.29\pm0.47$ & $0.79\pm0.27$ & WDMS & 2.48 \\
0218+0057 & 02:18:49.98 & +00:57:39.2 & 20.23 & 19.67 & 19.15 & $12250\pm1007$ & $7.46\pm0.59$ & $0.35\pm0.29$ & UNKN & 2.63 \\
0255-0044 & 02:55:09.29 & -00:44:14.7 & 20.76 & 19.69 & 18.74 & $13182\pm1916$ & $7.41\pm0.61$ & $0.36\pm0.27$ & WDMS & 2.71 \\
0327-0022 & 03:27:58.15 & -00:22:15.4 & 20.47 & 19.48 & 18.63 & $12902\pm874 $ & $7.80\pm0.31$ & $0.50\pm0.17$ & WDMS & 3.17 \\
0328+0017 & 03:28:42.92 & +00:17:49.7 & 19.06 & 18.03 & 16.16$^{2}$ & $12504\pm514 $ & $7.56\pm0.15$ & $0.40\pm0.07$ & WDMS & 2.39 \\ 
0336-0047 & 03:36:56.12 & -00:47:27.8 & 20.09 & 19.49 & 19.23 & $15246\pm1042$ & $7.79\pm0.24$ & $0.50\pm0.13$ & WDMS & 3.10 \\
0345-0614 & 03:45:14.71 & -06:14:21.2 & 20.16 & 19.31 & 18.51 & $13904\pm1267$ & $8.24\pm0.27$ & $0.76\pm0.17$ & PCEB & 3.61 \\
0824+1723$^{3}$ & 08:24:29.02 & +17:23:45.4 & 19.06 & 18.34 & 17.83 & $11433\pm358 $ & $8.21\pm0.18$ & $0.74\pm0.12$ & WDMS & 2.52 \\ 
1043+0603$^{4}$ & 10:43:58.59 & +06:03:20.9 & 19.24 & 18.75 & 18.78 & $11173\pm299 $ & $8.19\pm0.20$ & $0.72\pm0.13$ & WDMS & 2.73 \\ 
1054+1008 & 10:54:36.18 & +10:08:37.3 & 18.98 & 18.61 & 18.58 & $11433\pm229 $ & $7.97\pm0.13$ & $0.59\pm0.09$ & UNKN & 1.87 \\
1117-1255 & 11:17:10.54 & -12:55:40.9 & 20.30 & 19.65 & 19.25 & $11302\pm357 $ & $8.29\pm0.24$ & $0.79\pm0.15$ & WDMS & 1.18 \\
1136+0409 & 11:36:55.17 & +04:09:52.6 & 17.57 & 17.07 & 17.17 & $11699\pm152 $ & $7.99\pm0.08$ & $0.60\pm0.05$ & PCEB & 1.05 \\
1223-0056 & 12:23:39.61 & -00:56:31.2 & 18.41 & 17.91 & 18.06 & $11565\pm159 $ & $7.71\pm0.11$ & $0.45\pm0.06$ & PCEB$^{5}$ & 7.75$^{6}$ \\
1228-0225 & 12:28:50.46 & -02:25:09.4 & 20.74 & 19.61 & 18.52 & $13127\pm6398$ & $8.96\pm0.73$ & $1.18\pm0.38$ & WDMS & 1.62 \\
1329+2557 & 13:29:33.67 & +25:57:43.2 & 20.09 & 18.75 & 17.75 & $11565\pm1177$ & $8.43\pm0.44$ & $0.88\pm0.26$ & UNKN & 0.71 \\
1453+0010 & 14:53:05.77 & +00:10:48.2 & 19.71 & 18.90 & 18.21 & $11565\pm233 $ & $8.55\pm0.09$ & $0.95\pm0.05$ & WDMS & 1.93 \\
1520+0634 & 15:20:33.43 & +06:34:42.9 & 19.47 & 18.91 & 18.52 & $11302\pm234 $ & $8.47\pm0.12$ & $0.86\pm0.09$ & WDMS & 1.39 \\
1615+2357 & 16:15:05.51 & +23:57:46.3 & 19.02 & 18.23 & 17.66 & $12250\pm595 $ & $8.11\pm0.12$ & $0.67\pm0.08$ & UNKN & 1.99 \\
1652+1340 & 16:52:40.74 & +13:40:15.0 & 20.95 & 19.98 & 19.11 & $11173\pm101 $ & $8.44\pm0.44$ & $0.88\pm0.25$ & UNKN & 1.31 \\
1724+0733 & 17:24:45.28 & +07:33:24.7 & 19.51 & 18.86 & 18.35 & $13588\pm527 $ & $8.02\pm0.12$ & $0.62\pm0.08$ & WDMS & 1.85 \\
\hline
\multicolumn{11}{l}{$^{*}$ These magnitudes are on the SDSS 2.5\,m photometric system. Note, however, that ULTRACAM filters are primed,} \\
\multicolumn{11}{l}{\phantom{$^{*}$ }i.e. $u\,', g\,', r\,', i\,'$, as they are closer to the USNO 40-in system \citep{fukugitaetal96-1,smithetal02-1}.} \\
\multicolumn{11}{l}{$^{a,b,c}$ Values obtained using the updated spectral decomposition/fitting technique described in detail in 
\citet{rebassamansergasetal12-1}.} \\
\multicolumn{11}{l}{$^{1}$Total time over two observing nights ; $^{2}$This is an $i$ magnitude, and the system was observed in the $i\,'$-band ; $^{3}$Exposure time for} \\
\multicolumn{11}{l}{ $u\,'$-band was 40 sec ; $^{4}$Exposure time for all bands was 10 sec ; $^{5}$Originally mis-classified as a WDMS, see Sec.\,\ref{subsec:ecli} ; $^{6}$Total time} \\
\end{tabular}
\end{table*}

The class of ZZ\,Ceti stars comprises single, hydrogen-atmosphere (DA), photometrically variable white dwarfs (WDs) 
\citep[see e.g.][]{mcgraw77-1,clemens93-1,mukadametal06-1} that exhibit non-radial g-mode pulsations 
\citep{chanmugam72-1,warner+robinson72-1}. ZZ\,Ceti stars are tightly grouped together in the \textit{ZZ\,Ceti instability strip} \citep[e.g.][]{bergeronetal95-1,koester+allard00-1,mukadametal04-2,gianninasetal06-1,vangrooteletal12-1}, a well defined region in the effective temperature (\twd)-surface gravity (\sugr) plane, between 11,100 and 12,600 K for a $\sugr\,=\,8$ WD \citep{gianninasetal11-1}, with a dependence on \sugr\, \citep{giovanninietal98-1}. One of the fundamental questions regarding the
instability strip is its assumed purity \citep[e.g.][]{gianninasetal05-1,gianninasetal06-1,castanheiraetal10-1}. While available evidence points to a pure strip \citep[e.g.][]{bergeronetal95-1,bergeronetal04-1,castanheiraetal07-1}, the purity has not yet been explicitly proved. A pure strip implies that ZZ\,Ceti-type pulsations are an evolutionary stage of all (single) DA white dwarfs\footnote{An impure strip doesn't necessarily imply the opposite. Impurities could arise if a third parameter is in effect, as is the case for cataclysmic variables, explained later in this Section. In such a case, the third parameter needs to be taken into account when defining the strip.} and thus, studies of ZZ\,Cetis can be applied to the entire population of DA white dwarfs. The means towards this end is asteroseismology.

Asteroseismology, the study of stellar pulsations, is a powerful tool to gather direct information about the interiors of stars. In the past few decades it has been successfully adapted to WDs to constrain the hydrogen layer mass, the depth of the degenerate core boundary, the overall stellar mass and temperature, the mass of any other chemically stratified layers, and the rotation rates of these stellar remnants \citep{winget+kepler08-1,fontbrass08-1,althausetal10-1}.

As the end-point of all low-mass stars, WDs are also frequently found in binary configurations, such as cataclysmic variables (CVs), close interacting binaries in which the WD primary accretes from a low-mass companion. With the discovery of ZZ\,Ceti-type pulsations in the WD primary of the CV GW\,Lib \citep{warner+vanzyl98-1,vanzyletal00-1,vanzyletal04-1}, the analytic power of asteroseismology has become available to determine accurate stellar parameters for the WDs in these binaries, which is otherwise very difficult, if not impossible, to achieve, as the light from the WD is contaminated with emission from the accretion disc. Asteroseismology also offers the opportunity to study how the accretion of mass and angular momentum onto the WDs in CVs can affect the WD structure, shedding light on the potential single-degenerate scenario for SN Type-Ia progenitors. Most of the (non-magnetic) WDs in CVs were not expected to be pulsating, as accretion heats them to effective temperatures greater than $12000\,\mathrm{K}$ \citep{townsley+gaensicke09-1}. The work of \citet{townsleyetal04-1} and \citet{arrasetal06-1} suggested that the atmospheric composition of accreting WDs, especially the He abundance, the accretion of heavier elements and the rapid rotation of the WD primaries in CVs could significantly affect the pulsation modes. There are currently 16 known pulsating WDs in CVs. Most of them show a single pulsation mode, and thus it has not been possible to derive stellar parameters through asteroseismology for any of them. Their corresponding instability strip seems to be wider than the one for non-interacting ZZ\,Cetis, ranging from $10500\,\mathrm{K}\,\lesssim\,T_{\mathrm{eff}}\,\lesssim\,16000\,\mathrm{K}$ \citep[][and references therein]{szkodyetal10-1}, although this result is based on small-number statistics. However, its location as derived from observations is consistent with theoretical expectations \citep[e.g.][]{arrasetal06-1}, for a high He abudance ($>$ 0.48). We should note that the CV strip is not pure, most likely a consequence of the 3-dimensional parameter space, with the He abundance as the third parameter, in addition to \twd\, and \sugr. 

The occurrence of pulsating WDs in CVs raises the question of the existence of pulsating WDs in the progenitors of CVs and, generally, in non-interacting binaries, such as detached white dwarf plus main-sequence binaries (WD+MS). In recent years, owing mainly to the Sloan Digital Sky Survey (SDSS, \citealt{yorketal00-1}), WD+MS binaries have been discovered in large numbers \citep[e.g.][]{silvestrietal05-1,helleretal09-1,rebassamansergasetal10-1,rebassamansergasetal12-1,liuetal12-1,rebassamansergasetal13-1}, offering the exciting possibility for ensemble asteroseismology studies of a large and homogeneous sample of WDs in binaries.

About two-thirds of the SDSS WD+MS binaries are wide enough that the WD progenitors have evolved as if they were single stars 
\citep[e.g.][]{willems+kolb04-1,schreiberetal10-1,nebotetal11-1}. Thus, pulsating WDs in such systems should exhibit properties very similar to the single ZZ\,Cetis. This hypothesis remains effectively untested by observations; to our knowledge, there is only one confirmed ZZ\,Ceti in a wide common proper motion binary, G117-B15A \citep[e.g.][]{kepleretal91-1}.

The rest of the WD+MS binaries are expected to be post-common envelope binaries (PCEBs), i.e. close binaries that have formed through common envelope (CE) evolution \citep[e.g.][for a review]{webbink08-1}. The CE phase is believed to be the main mechanism for the formation of low mass, He-core WDs \citep[e.g][]{paczynski76-1,iben+tutukov86-1}.

While the presence of such WDs in PCEBs has been established observationally, in both double-degenerate (\citealt{marshetal95-1,marsh95-1,steinfadtetal10-1,parsonsetal11-1}; and see also \citealt{nelemans+tout05-1} and references therein) and single-degenerate (i.e. with a MS companion) configurations (\citealt{marshduck96-1,bruch99-1,pyrzasetal12-1,parsonsetal12-1} and see also \citealt{schreiber+gaensicke03-1} and references therein), pulsating He-core WDs have resisted detection \citep{steinfadtetal12-1} and have been discovered only very recently, in double-degenerate configurations \citep[extremely low mass WDs (ELM),][]{hermesetal12-1,hermesetal13-1,hermesetal13-2}. Previous to our work, no pulsating WDs in single-degenerate PCEBs were known.

The study of pulsating WDs in PCEBs can provide crucial insight into whether, and if so how, the internal structure of WDs, and by extension the pulsation properties, are affected by the common envelope phase. White dwarfs in single-degenerate PCEBs  can be found mainly in two ``flavours'', those with $\mwd\,<\,0.45\,\Msun$ and a He-core and those with $\mwd\,\ge\,0.5\,\Msun$ and a C/O core \citep[e.g.][]{pyrzasetal09-1,rebassamansergasetal11-1}; the juxtaposition of the two classes can provide fertile ground for studies on the potential effects of mass loss on the WD structure and pulsations. Furthermore, these systems can populate the gap in WD mass between the normal ZZ Cetis (with $\mwd\,=\,0.6\,\Msun$ and above) and the ELMs (with $\mwd\,<\,0.25\,\Msun$), leading to a complete census of pulsating WDs.

In this paper we present the first results from a dedicated search for pulsating WDs in WD+MS binaries. The structure is as follows: our target selection, photometric observations and period analysis techniques are described in Sec.\,\ref{sec:tarsel} and \ref{sec:obsnred} respectively, while Section\,\ref{sec:res} presents our results. We discuss our findings in Sec.\,\ref{sec:disc} and conclude in Sec.\,\ref{sec:concl} with a summary and suggestions for future work.

\begin{table*}
\setlength{\tabcolsep}{0.7ex}
\centering
\caption{Summary of the pulsation properties of the seven ZZ Ceti white dwarfs. We provide the frequency, the corresponding period and the amplitude of each detection for all three ULTRACAM arms, as well as the corresponding detection thresholds. A dash indicates no detection.}
\label{tab:respulssys}
\begin{tabular}{@{}ccccccccccccc@{}}
\hline 
System & \multicolumn{3}{c}{Frequency [$\mu$Hz]} & \multicolumn{3}{c}{Period [s]} & \multicolumn{3}{c}{Amplitude [mma$^{1}$]} & \multicolumn{3}{c}{Det. Thresh. [mma]} \\
SDSS\,J & $u\,'$ & $g\,'$ & $r\,'$ & $u\,'$ & $g\,'$ & $r\,'$ & $u\,'$ & $g\,'$ & $r\,'$ & $u\,'$ & $g\,'$ & $r\,'$ \\
\hline 
0052-0051$^{2}$ & - & $928.36\,\pm\,12.73$ & $932.18\,\pm\,9.84$ & - & $1077.2\,\pm\,14.8$ & $1072.8\,\pm\,11.3$ & - & 4.0 & 2.4 & 30.8 & 2.7 & 2.2 \\
 & - & $914.24\,\pm\,5.21$ & - & - & $1093.8\,\pm\,6.2$ & - & - & 4.0 & - & 32.3 & 2.5 & 2.2 \\
0111+0009 & $1130.94\,\pm\,8.30$ & $1131.68\,\pm\,8.15$ & $1134.34\,\pm\,17.55$ & $884.2\,\pm\,6.5$ & $883.6\,\pm\,6.4$ & $881.6\,\pm\,13.6$ & 33.2 & 15.9 & 6.2 & 22.0 & 9.0 & 4.5 \\
 & $1579.63\,\pm\,7.87$ & $1583.34\,\pm\,4.51$* & $1581.13\,\pm\,5.21$ & $633.1\,\pm\,3.2$ & $631.6\,\pm\,1.8$ & $632.5\,\pm\,2.1$ & 40.2 & 28.0 & 10.1 & 22.0 & 9.0 & 4.5 \\
 & $1715.74\,\pm\,9.49$ & $1714.58\,\pm\,5.21$ & $1704.75\,\pm\,7.06$ & $582.8\,\pm\,3.2$ & $583.2\,\pm\,1.8$ & $586.6\,\pm\,2.4$ & 21.4 & 16.3 & 6.3 & 22.0 & 9.0 & 4.5 \\
 & $1950.12\,\pm\,8.45$ & $1960.07\,\pm\,5.44$ & $1955.56\,\pm\,8.10$ & $512.8\,\pm\,2.2$ & $510.2\,\pm\,1.4$ & $511.4\,\pm\,2.1$ & 25.5 & 18.9 & 7.7 & 22.0 & 9.0 & 4.5 \\
 & - & $2728.59\,\pm\,14.47$ & $2723.26\,\pm\,12.85^{a}$ & - & $366.5\,\pm\,1.9$ & $367.2\,\pm\,1.7$ & - & 9.1 & 4.4 & 22.0 & 9.0 & 4.5 \\
0203+0040 & $1039.24\,\pm\,7.87$ & $1044.91\,\pm\,2.78$* & $1049.77\,\pm\,4.40$ & $962.4\,\pm\,7.3$ & $957.0\,\pm\,2.6$ & $952.6\,\pm\,4.0$ & 70.8 & 38.5 & 11.7 & 46.2 & 8.5 & 4.7 \\
 & $1456.48\,\pm\,10.19^{a}$ & $1462.26\,\pm\,7.41$ & $1483.91\,\pm\,9.14$ & $685.9\,\pm\,4.8$ & $683.9\,\pm\,3.5$ & $673.9\,\pm\,4.2$ & 41.8 & 15.2 & 4.8 & 46.2 & 8.5 & 4.7 \\
 & - & $2507.75\,\pm\,8.56$ & $2510.30\,\pm\,11.69$ & - & $398.8\,\pm\,1.4$ & $398.4\,\pm\,1.9$ & - & 11.9 & 4.7 & 46.2 & 8.5 & 4.7 \\
0824+1723 & $1010.88\,\pm\,11.34$ & $1012.26\,\pm\,8.91$ & $1031.48\,\pm\,7.18$ & $989.2\,\pm\,11.1$ & $987.9\,\pm\,8.7$ & $969.5\,\pm\,6.7$ & 27.3 & 12.0 & 6.7 & 25.1 & 7.5 & 4.3 \\
 & $1178.82\,\pm\,14.58$ & $1238.88\,\pm\,10.30$ & $1190.16\,\pm\,11.92$ & $848.3\,\pm\,10.5$ & $807.2\,\pm\,6.7$ & $840.2\,\pm\,8.4$ & 29.2 & 13.9 & 5.7 & 25.1 & 7.5 & 4.3 \\
 & $1313.77\,\pm\,17.82$ & -$^b$ & $1300.12\,\pm\,19.33$ & $761.2\,\pm\,10.3$ & - & $769.2\,\pm\,11.4$ & 28.6 & - & 4.6 & 25.1 & 7.5 & 4.3 \\
 & $1588.43\,\pm\,14.35$ & $1603.47\,\pm\,4.75$* & $1626.97\,\pm\,11.69$ & $629.6\,\pm\,5.7$ & $623.7\,\pm\,1.9$ & $614.7\,\pm\,4.4$ & 30.4 & 20.4 & 7.5 & 25.1 & 7.5 & 4.3 \\
 & - & $1945.60\,\pm\,9.83$ & $1967.36\,\pm\,11.11$ & - & $513.9\,\pm\,2.6$ & $508.3\,\pm\,2.9$ & - & 9.2 & 4.7 & 28.0 & 7.5 & 4.3 \\
1043+0603 & - & $1576.27\,\pm\,6.71$ & $1567.71\,\pm\,11.69$ & - & $634.4\,\pm\,2.7$ & $637.9\,\pm\,4.8$ & - & 8.5 & 7.1 & 24.6 & 4.5 & 5.8 \\
 & - & $3080.21\,\pm\,2.20$* & $3082.06\,\pm\,3.70$ & - & $324.7\,\pm\,0.2$ & $324.6\,\pm\,0.4$ & - & 28.8 & 18.3 & 24.6 & 4.5 & 5.8 \\
 & - & $3419.10\,\pm\,15.97$ & - & - & $292.5\,\pm\,1.4$ & - & - & 5.3 & - & 24.6 & 4.5 & 5.8 \\
 & - & $6167.94\,\pm\,7.52^{\dagger}$ & $6143.52\,\pm\,18.63^{a}$ & - & $164.9\,\pm\,0.2$ & $162.8\,\pm\,0.5$ & - & 6.6 & 5.5 & 24.6 & 4.5 & 5.8 \\
1117-1255 & - & $1196.53\,\pm\,15.51$* & $1248.26\,\pm\,18.87^{a}$ & - & $835.80\,\pm\,10.80$ & $801.10\,\pm\,12.10$ & - & 21.9 & 13.1 & 77.2 & 10.9 & 13.6 \\
1136+0409 & - & $3616.67\,\pm\,13.31$* & $3656.37\,\pm\,17.36$ & - & $276.5\,\pm\,1.0$ & $273.5\,\pm\,1.3$ & - & 8.3 & 6.3 & 22.1 & 3.6 & 4.9 \\
 & - & $5489.70\,\pm\,19.91$ & - & - & $182.2\,\pm\,0.7$ & - & - & 4.4 & - & 22.1 & 3.6 & 4.9 \\
\hline
\multicolumn{13}{l}{$^{1}$ mma: milli-modulation amplitude, a 0.1\% relative amplitude change ; $^{2}$ Two observations - solution for each independent observation} \\
\multicolumn{13}{l}{* Pre-whitened signal ; $^{a}$ Only $>\,2\,\sigma$ detection ; $^{b}$ Barely $	1\,\sigma$ detection ; $^{\dagger}$ First harmonic of the 3080.21 signal} \\
\end{tabular}
\end{table*}


\section{Target selection}
\label{sec:tarsel}

Targets were selected from the white dwarf plus main-sequence binaries catalogue by \citet{rebassamansergasetal12-1}. Systems were chosen on the basis of their \sugr\, and \twd\, values, as determined from their SDSS spectra, using the spectral decomposition/fitting technique described in detail in \citet{rebassamansergasetal07-1} 
\citep[and see also][]{rebassamansergasetal12-1}. Table\,\ref{tab:allinfo} lists basic information on all our targets and the observations. Based on available radial velocity (RV) information, either from the original SDSS spectroscopy or from follow-up observations, these 26 targets can be divided into three groups:

\begin{enumerate}
 \item PCEB: systems for which spectroscopy reveals radial velocity variations (close binaries)
 \item WDMS: systems for which spectroscopy reveals no radial velocity variations (wide binary candidates)
 \item UNKN: systems for which no or insufficient spectroscopy is available
\end{enumerate}

While the detection of short period RV variations unambiguously identifies the PCEBs among our targets, the absence of RV variations is not a guarantee that a system is a wide binary. Insufficient spectral resolution, low orbital inclination or an unlucky orbital phase sampling can prevent the identification of a PCEB \citep[see e.g.][for a discussion]{schreiberetal10-1}. An example is SDSS\,J1223-0056, whose available spectroscopy did not show any RV variations and the system was classified as a WDMS. However, our subsequent photometric observations revealed the system to be eclipsing (see Sec.\,\ref{subsec:ecli}); the orbital period was determined to be $P_{\mathrm{orb}}\,=\,2.2\,$h, unambiguously re-classifying the system as a PCEB. Thus, WDMS should be read as ``candidate wide WD+MS binary.''


\section{Observations, Reductions and Analysis}
\label{sec:obsnred}

\subsection{Photometry: NTT/ULTRACAM}
\label{subsec:photom}

Photometric light curves of our targets were obtained during two observing runs in December 2010 and May 2011. We used the high-speed camera ULTRACAM \citep{dhillonetal07-1} mounted as a visitor instrument on ESO's 3.5\,m New Technology Telescope (NTT) at La Silla Observatory, Chile. Each target was observed in full-frame mode, for a continuous block of time, but varying in length depending on the schedule. The exposure time was 20\,s, with a dead-time between exposures of $\sim\,25\,$ms. ULTRACAM is a triple-beam camera, so data were obtained simultaneously in the Sloan $u\,'$, $g\,'$ and $r\,'$ bands. In one case an $i\,'$ filter was used instead of $r\,'$ for scheduling reasons. All of the data were reduced with aperture photometry using the ULTRACAM pipeline software, with debiassing, flat-fielding and sky background subtraction performed in the standard way. The fluxes of the targets were determined using a variable aperture, whereby the radius of the aperture is scaled according to the full width at half-maximum of the stellar profile. Variations in transparency were accounted for by dividing each light curve by the light curve of a nearby comparison star. The stability of these comparison stars was checked against other stars in the field, and no variations were seen.


\begin{figure*}
\centering
 \includegraphics[angle=-90,width=\textwidth]{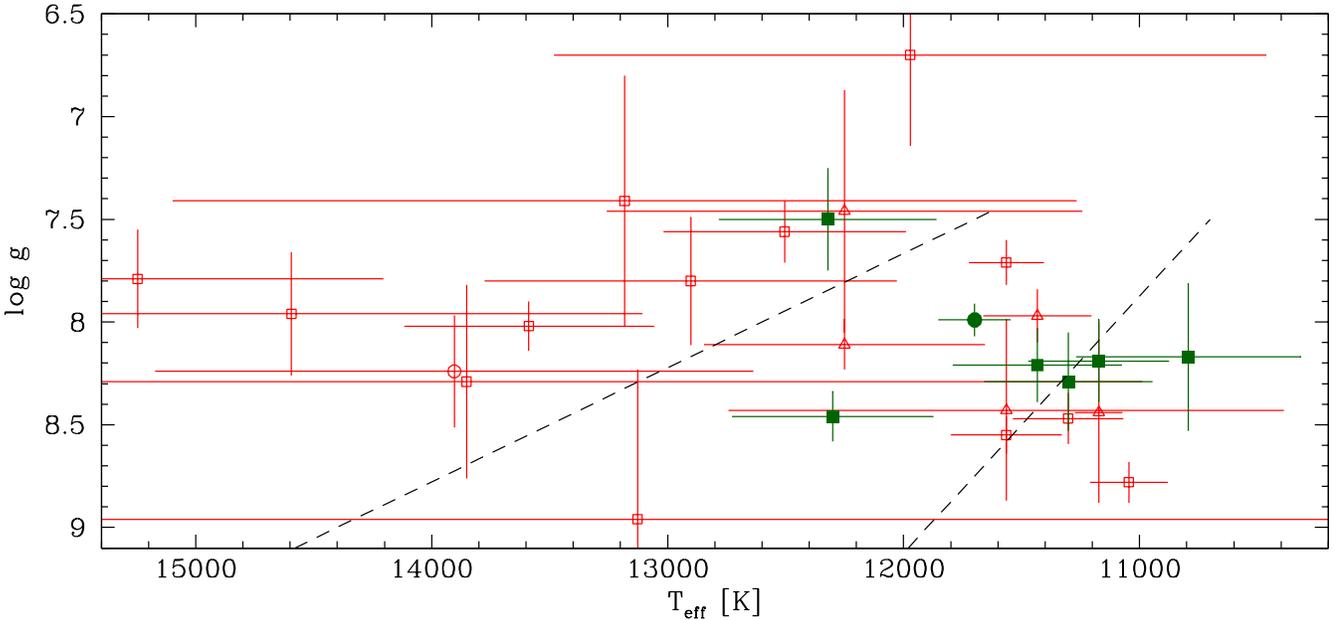}
  \caption{The results of our campaign. The 26 systems plotted on the \twd\,-\,\sugr\, plane. Pulsating systems are shown with green, filled symbols; non-pulsating systems with red, open symbols (colour version available only online). Circles denote PCEB, squares WDMS and triangles UNKN binaries, following Table\,\ref{tab:allinfo}. The dashed lines are the boundaries of the ZZ\,Ceti instability strip found by \citet{gianninasetal11-1}.}
  \label{fig:results}
\end{figure*}

\begin{figure*}
\centering
 \includegraphics[angle=-90,width=\textwidth]{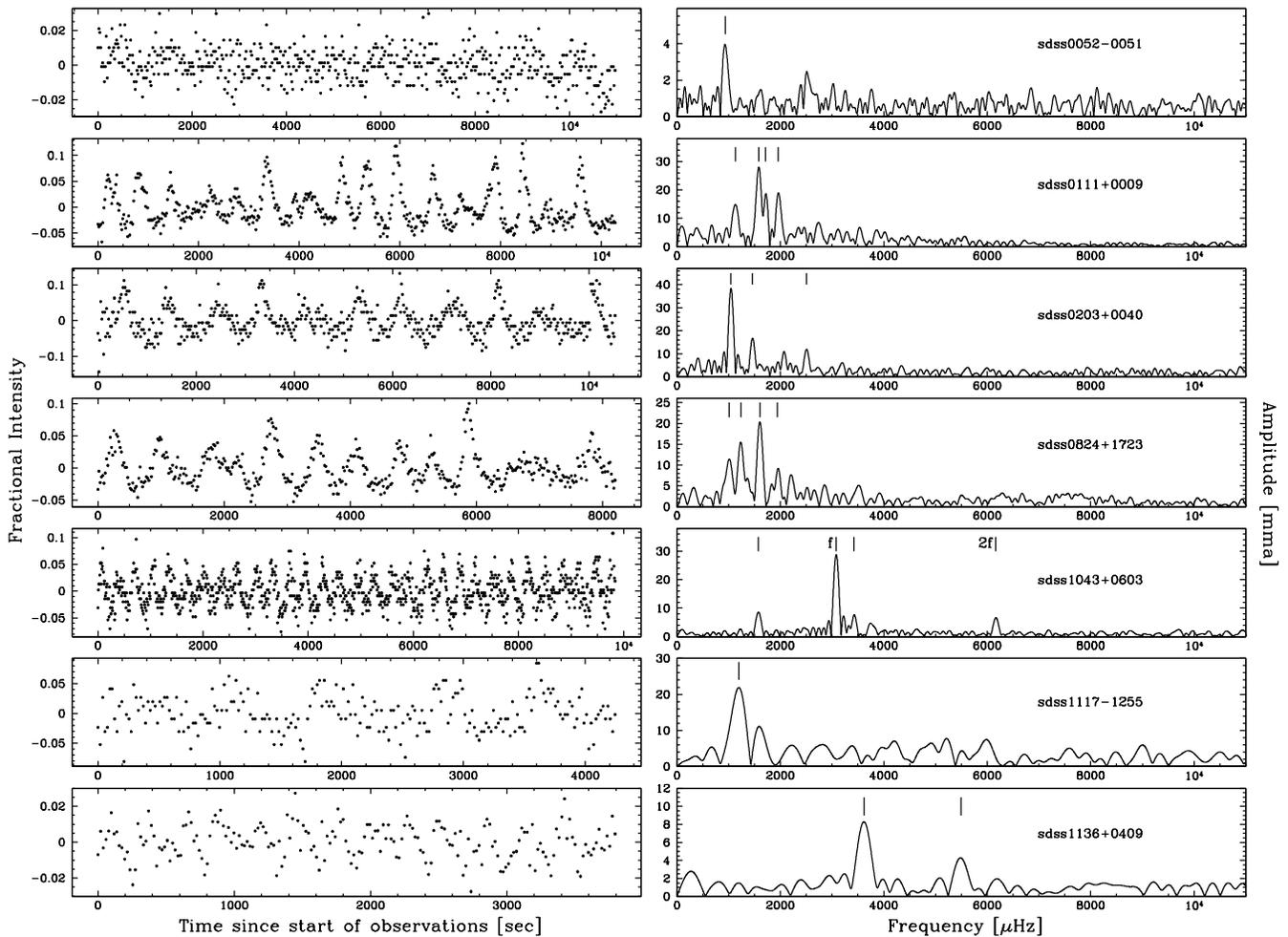}
  \caption{The seven pulsating WDs: $g\,'$-band light curves (left panels) and the corresponding amplitude spectra (right panels) for each system, identified in the right panels. Tickmarks indicate the significant detections for each system, as listed in Table\,\ref{tab:respulssys}.}
  \label{fig:samplelc}
\end{figure*}

\subsection{Period analysis}
\label{subsec:dftan}

For the period analysis and the detection of pulsations, each light curve was first converted into fractional intensity (dividing by the mean and subtracting 1). Subsequently, we calculated amplitude spectra using the {\sc TSA} package \citep{schwarzenbergczerny93-1} as implemented in \texttt{MIDAS}. 

In order to judge the significance of a signal, we calculated a detection threshold in the following fashion \citep[see also][]{greissetal14-1}: for each light curve we created artificial data sets using a shuffling technique \citep[see e.g.][for discussions]{keplerso93-1,schreiber07-1}, where each intensity point $f_{i}$ is randomly re-assigned to a time point $t_{j}$ with $i\neq j$; all intensity and time points are used. This shuffling destroys any coherent signal, but retains the time- and overall noise properties of a light curve. We then calculated the amplitude spectra of the shuffled light curves and recorded the value of highest amplitude. Using the results from 10,000 shufflings, we calculated the amplitudes corresponding to the 68.3, 95.5 and 99.7 per cent confidence levels (1-, 2- and 3-$\sigma$ respectively) and set the 3\,$\sigma$ amplitude as our detection threshold. Signals with amplitudes above this threshold were considered significant. Finally, the frequencies and errors of significant detections were determined using the bootstrap method \citep{press02-1}.

When strong signals are present, using the maximum amplitude to determine the threshold typically leads to overestimated threshold values. Therefore, our analysis was carried out as follows: for each light curve, we used our shuffling technique to calculate the detection threshold and to ensure that the strongest signal present in the power spectrum is indeed a 3-$\sigma$ detection. Subsequently, we prewhitened this signal, and subjected the light curve to a second shuffling, in order to calculate the revised detection threshold and look for additional significant signals. We did not proceed with further iterations, as we have only one relatively short light curve per target.


\section{Results}
\label{sec:res}

\begin{table}
\centering
\caption{Detection thresholds for the 19 white dwarfs where no pulsations were detected.}
\label{tab:resnonpulssys}
\begin{tabular}{@{}cccc@{}}
\hline 
System & \multicolumn{3}{c}{Threshold [mma]} \\
SDSS\,J & $u\,'$ & $g\,'$ & $r\,'$ \\
\hline 
0017-0024 & 21.3 & 4.4 & 5.5 \\
0021-1103 & 14.4 & 3.3 & 3.4 \\
0124-0023 & 38.5 & 4.2 & 2.5 \\ 
0212+0018 & 19.9 & 4.1 & 4.5 \\
0218+0057 & 94.6 & 21.7 & 10.9 \\
0255-0044 & 62.6 & 10.2 & 4.8 \\
0327-0022 & 72.4 & 9.6 & 6.7 \\
0328+0017 & 51.7 & 5.2 & 1.2 \\
0336-0047 & 30.2 & 6.3 & 5.67 \\
0345-0614 & 54.3 & 8.2 & 4.2 \\
1054+1008 & 50.2 & 10.1 & 15.5 \\
1223-0056 &  9.7 & 2.1 & 3.1 \\
1228-0225 & 217.2 & 26.0 & 13.9 \\
1329+2557 & 87.0 & 8.3 & 6.2 \\
1453+0010 & 25.6 & 4.1 & 3.3 \\
1520+0634 & 22.5 & 4.6 & 5.1 \\
1615+2357 & 33.5 & 4.3 & 4.6 \\
1652+1340 & 92.6 & 13.8 & 9.3 \\
1724+0733 & 36.5 & 6.3 & 7.4 \\
\hline
\end{tabular}
\end{table}

\begin{figure*}
\centering
 \includegraphics[angle=-90,width=\textwidth]{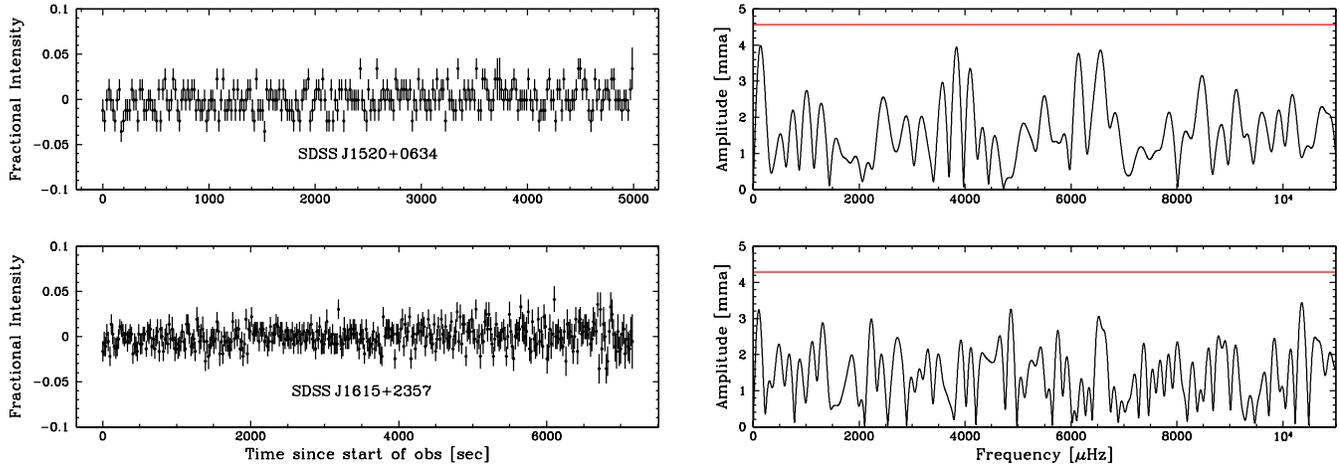}
  \caption{Sample $g\,'$-band light curves (left panels) and their corresponding amplitude spectra (right panels) of two of our non-pulsating WDs, SDSS\,J1520+0634 (top panels) and SDSS\,J1615+2357 (bottom panels). Red lines indicate the 3\,$\sigma$ detection threshold.}
  \label{fig:samplelcnop}
\end{figure*}

Our photometric observations and subsequent period analysis revealed the pulsating nature of seven white dwarfs in our sample, while the remaining 19 were found to be non-variable at our detection threshold. Our results are illustrated in Figure\,\ref{fig:results}.

Table\,\ref{tab:respulssys} lists the seven systems with $>\,3\,\sigma$ signal detections, along with the corresponding frequencies, periods, amplitudes and detection thresholds in each filter, while Fig.\,\ref{fig:samplelc} shows the $g\,'$-band light curves and the corresponding amplitude spectra of all seven pulsating systems. 

In Table\,\ref{tab:resnonpulssys} we list all the non-pulsating systems along with the corresponding detection thresholds in each of the three filters, while Fig.\,\ref{fig:samplelcnop} shows two sample $g\,'$-band light curves and their respective amplitude spectra.

The periods observed in our systems are comparable to those observed in typical pulsating WDs \citep[e.g.][]{mukadametal06-1}. All our systems are detached, so we do not expect to see any accretion-related variability, such as  flickering and QPO's. In detached systems, photometric variability could be caused by ellipsoidal modulation or irradiation/reflection effects \citep[see e.g.][]{pyrzasetal09-1,parsonsetal10-1}. However, these forms of variability are modulated on longer timescales (effectively on the orbital period, with two maxima and one maximum per orbit respectively) than the periods observed in our sample, and have a distinctive qualitative imprint on the light curves, very different to the observed variability. Henceforth, we will \emph{assume} that each $>\,3\,\sigma$ detection corresponds to a pulsation mode, although repeated detections of each pulsation would be required for unambiguous confirmation.

\subsection{SDSS\,J1223-0056}
\label{subsec:ecli}

\begin{figure}
\centering
 \includegraphics[angle=-90,width=84mm]{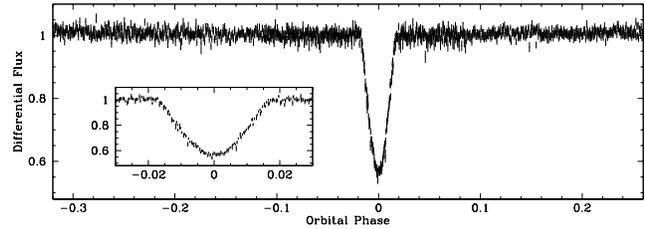}
  \caption{Phase-folded $g\,'$-band lightcurve of SDSS\,J1223-0056, using the orbital ephemeris Eq.\,\ref{eq:ephe}. Inset panel: a zoom-in version around the eclipse itself.}
  \label{fig:ecli}
\end{figure}

Our photometric observations revealed that the white dwarf in SDSS\,J1223-0056 undergoes partial eclipses. Measuring the times of mid-eclipse from four observed eclipses, we determine the orbital ephemeris of the system to be

\begin{equation}
\mathrm{BMJD_{0}(TDB)}\,=55706.11231(1)\,+\,E*0.0900844(9)
\label{eq:ephe}
\end{equation}

\noindent
that is $P_{\mathrm{orb}}\,=\,2.16202(2)\,$[h], where the numbers in brackets indicate the error on the last digit. Figure\,\ref{fig:ecli} shows a phase-folded $g\,'$-band lightcurve of the new eclipsing system. SDSS\,J1223-0056 has also been detected as an eclipsing system in data from the Catalina Real-time Transient Survey \citep[see][]{parsonsetal13-1}.

\section{Discussion}
\label{sec:disc}

Inspection of Tab.\,\ref{tab:respulssys} and \ref{tab:resnonpulssys} reveals that in the majority of cases, our $u\,'$-band detection threshold is much higher than the $g\,'$- and $r\,'$-band, since the quality of the data in this band is compromised by the faintness of the targets and the lack of good comparison stars in the fields. As a result, in some of the pulsating systems there is no $u\,'$-band detection of the pulsations, although the respective $g\,'$- and $r\,'$-band detections are unambiguous. 

Our survey for pulsations generally has a relatively low sensitivity; the median value for WDs not observed to vary is 6.3 mma in the $g\,'$-band, only slightly below the median amplitude of 8.8 mma for detected pulsations in a sample of 35 ZZ Cetis discovered from SDSS \citep{mukadametal04-1}. Pulsation signals could also be diluted if the M-dwarf companion contributes a significant amount of flux. Thus, it is possible that we might have missed some pulsators among these seemingly non-variable systems. 

For our seven pulsating systems, the frequencies of the detected signals are consistent with those of g-mode pulsations observed in single WDs \citep[e.g.][]{mukadametal06-1}. The multicolour amplitudes of the pulsations (particularly in the cases where $u\,'$-band pulsations are also detected) are indicative of $\ell\,=\,1,2$ modes \citep{robinsonetal95-1}. Three systems (SDSS\,J0111+0009, J0824+1723 and J1043+0603, all of them wide WDMS candidates) show multiple pulsation periods, making them promising candidates for intensive follow-up asteroseismic studies. Among the three confirmed PCEBs in our sample, only SDSS\,J1136+0409 is pulsating, while J1223-0056 and J0345-0614 show no pulsations at our detection limit.

Finally, we note that if we take the \twd\, and \sugr\, of these WDs at face value, two of our pulsating systems (SDSS\,J0111+0009 and J0203+0040) lie nominally outside the \citet{gianninasetal11-1} instability strip, while six systems with no detections lie inside it. However, it is important to bear in mind that the parameters determined from the spectral decomposition \citep{rebassamansergasetal10-1,rebassamansergasetal12-1} have rather large statistical errors (evident in the errors bars of Fig.\,\ref{fig:results}) and can be subject to systematic uncertainties \citep[see][]{parsonsetal13-1}. In WD+MS binaries, the subtraction of the M-dwarf component from the spectrum directly affects the disambiguation between the ``hot'' and ``cold'' solutions of the subsequent WD fit. It is possible that for some of our systems we have selected the ``wrong'' solution. Also, \sugr\, values are overestimated for those systems with $\twd\,<\,12000\,\mathrm{K}$, when using 1-dimensional WD models to perform the fits. Recently \citet{tremblayetal13-1} published 3-dimensional WD models, allowing for corrections to the 1D \sugr\, values. However, since these 3D corrections have not yet been implemented in the context of an empirical ZZ Ceti instability strip, we have not adopted them in our analysis.


\section{Conclusions}
\label{sec:concl}
We have carried out the first dedicated survey to identify pulsating white dwarfs in detached WD+MS binaries. Among a sample of 26 such systems, selected based on the \twd\, and \sugr\, values of their WDs obtained from spectral fitting, we have idenified seven new pulsating white dwarfs. One of these WDs is found in SDSS\,J1136+0409, a confirmed single-degenerate post-common-envelope binary,
which constitutes the first detection of such kind.

For the immediate future, work needs to be carried out on multiple fronts: (i) high signal-to-noise spectroscopy of all our targets with wide wavelentgh coverage (across the Balmer jump) in order to pinpoint their exact location on the \twd\,-\,\sugr\, plane, (ii) higher signal-to-noise photometric time series of the seemingly non-variable systems and (iii) follow-up photometric observations of all pulsating WDs in order to identify candidates for precision asteroseismology.


\section*{Acknowledgements}

We thank the referee Anjum Mukadam for a constructive report. Based on observations made with ESO telescopes at the La Silla Paranal Observatory under programme ID 086.D-0555 and 087.D-0557. The research leading to this results has received funding from the European Research Council under the European Union's Seventh Framework Programme (FP/2007-2013)/ERC Grant Agreement n.320964 (WDTracer). SP gratefully acknowledges support from an ALMA-CONICYT grant (31110019). ARM acknowledges financial support from the Postdoctoral Science Foundation of China (grant 2013M530470) and from the Research Fund for International Young Scientists by the National Natural Science Foundation of China (grant 11350110496). VSD, SPL and ULTRACAM are supported by the STFC. TRM and EB acknowledge the support of STFC under grant ST/L000733/1. SGP acknowledges support from FONDECYT in the form of grant number 3140585. MRS acknowledges support from FONDECYT (1141269). 


\bibliographystyle{mn_new} 
\bibliography{aamnem99,pulswdms}

\bsp

\label{lastpage}

\end{document}